\documentclass[english,aps,preprint]{revtex4}
\usepackage[T1]{fontenc}
\usepackage[latin9]{inputenc}
\usepackage{amsmath}
\usepackage{graphicx}
\usepackage{amssymb}
\usepackage{esint}

\makeatletter
\@ifundefined{textcolor}{}
{%
 \definecolor{BLACK}{gray}{0}
 \definecolor{WHITE}{gray}{1}
 \definecolor{RED}{rgb}{1,0,0}
 \definecolor{GREEN}{rgb}{0,1,0}
 \definecolor{BLUE}{rgb}{0,0,1}
 \definecolor{CYAN}{cmyk}{1,0,0,0}
 \definecolor{MAGENTA}{cmyk}{0,1,0,0}
 \definecolor{YELLOW}{cmyk}{0,0,1,0}
 }

\makeatother
\usepackage{babel}
\begin{document}

\title{Generating Transition Paths by Langevin Bridges}

\author{Henri Orland}

\email{henri.orland@cea.fr}

\affiliation{Institut de Physique Théorique, CEA, IPhT \\
 CNRS, URA2306, \\
 F-91191 Gif-sur-Yvette, France}
\begin{abstract}
We propose a novel stochastic method to generate paths conditioned
to start in an initial state and end in a given final state during
a certain time $t_{f}$. These paths are weighted with a probability
given by the overdamped Langevin dynamics. We show that these paths
can be exactly generated by a non-local stochastic differential equation.
In the limit of short times, we show that this complicated non-solvable
equation can be simplified into an approximate stochastic differential
equation. For longer times, the paths generated by this approximate
equation can be reweighted to generate the correct statistics. In
all cases, the paths generated by this equation are statistically
independent and provide a representative sample of transition paths. In case the reaction takes place in a solvent (e.g. protein folding in water), the explicit solvent can be treated.
The method is illustrated on the one-dimensional quartic oscillator. 
\end{abstract}
\maketitle

\section{Introduction}
The problem of finding the pathway of chemical or biological reactions
is of utmost importance for the understanding of their underlying
mechanisms, as it allows to have better control on these reactions
\cite{1}. For instance, in the realm of proteins, understanding the
pathway between the unfolded state and the native state, or between
two native states of the protein (allostery) may help prevent certain
reactions or on the contrary favor them. Recent progress in single
molecule experiments have allowed to monitor the spontaneous thermal
folding and unfolding of single proteins, or the force induced unfolding
of proteins \cite{3,4,5}.

In the following, we will study the spontaneous or the driven transition
between an initial state denoted A and a final state denoted B.

This problem has been addressed mainly by stochastic methods which
start from an initial path and deform it by sampling the vicinity
of the path. These are the path sampling methods \cite{6,7,8}. The
main drawback of these methods is that they are time consuming,
and they generate strongly correlated trajectories. As a consequence,
the space of sampled trajectories depends strongly on the initial used
path. The same kind of problem exists for the Dominant Pathway
method \cite{9,10}, where the minimal action path depends strongly
on the initial guess.

From now on, we assume that the system is driven by stochastic dynamics
in the form of an overdamped Langevin equation

\begin{equation}
\frac{dx}{dt}=-\frac{1}{\gamma}\frac{\partial U}{\partial x}+\eta(t)\label{eq:langevin}\end{equation}
 For the sake of simplicity, we illustrate the method on a one-dimensional
system, the generalization to higher dimensions or larger number of
degrees of freedom being straightforward. In this equation, $x(t)$
is the position of a point at time $t$ in a potential $U(x)$, $\gamma$
is the friction coefficient, related to the diffusion constant $D$
through the relation $D=k_{B}T/\gamma$, where $k_{B}$ is the Boltzmann
constant and $T$ the temperature of the thermostat. In addition,
$\eta(t)$ is a Gaussian white noise with moments given by

\begin{equation}
<\eta(t)>=0\label{eq:noise1}\end{equation}
 \begin{equation}
<\eta(t)\eta(t')>=\frac{2 k_{B}T}{\gamma}\delta(t-t')\label{eq:noise2}\end{equation}

It is well known that the probability distribution $P(x,t)$ for the
particle to be at point $x$ at time $t$ is given by a Fokker-Planck
equation \cite{11}

\begin{equation}
\frac{\partial P}{\partial t}=D\frac{\partial}{\partial x}\left(\frac{\partial P}{\partial x}+\beta\frac{\partial U}{\partial x}P\right)\label{eq:FP}\end{equation}
 where $\beta=1/k_{B}T$ is the inverse temperature. In this one dimensional
model, the initial state A is characterized by its position $x_{0}$
at time $0$ and the final state B by its position $x_{f}$ at time
$t_{f}$. This equation is thus to be supplemented by a boundary condition
$P(x,0)=\delta(x-x_{0})$ where $x_{0}$ is the initial position of
the particle.

It is convenient to go to the Schr\"odinger representation, by defining\[
\Psi(x,t)=e^{\beta U(x)/2}P(x,t)\]
The function $\Psi(x,t)$ satisfies the imaginary time Schr\"odinger
equation

\begin{equation}
\frac{\partial\Psi}{\partial t}=\frac{k_{B}T}{\gamma}\frac{\partial^{2}\Psi}{\partial x^{2}}-\frac{1}{4\gamma k_{B}T}V(x)\Psi(x)\label{eq:schrodinger}\end{equation}
 with \begin{equation}
V(x)=\left(\frac{\partial U}{\partial x}\right)^{2}-2k_{B}T\frac{\partial^{2}U}{\partial x^{2}}\label{eq:potential}\end{equation}
Using the standard notations of quantum mechanics, one can conveniently
write
\begin{equation}
P(x_{f},t_{f}|x_{0},0)=e^{-\beta(U(x_{f})-U(x_{0}))/2}<x_{f}|e^{-t_{f}H}|x_{0}>\label{eq:matrix}\end{equation}
where the Hamiltonian $H$ is given by

\begin{equation}
H=-\frac{k_{B}T}{\gamma}\frac{\partial^{2}}{\partial x^{2}}+\frac{1}{4\gamma k_{B}T}V(x)\label{eq:hamiltonian}
\end{equation}
In eq.(\ref{eq:matrix}), we have denoted by $P(x_{f},t_{f}|x_{0},0)$ the probability for a particle to start at $x_0$ at time $0$ and end at $x_f$ at time $t_f$, to emphasize the boundary conditions.

It is well-known that the ground state of $H$, which has $0$ energy,
is $\Psi_{0}(x)=e^{-\beta U(x)/2}/\sqrt{Z}$ where $Z$ is the partition
function of the system, and all eigenstates $\Psi_{\alpha}$ of $H$
have strictly positive energies $E_{\alpha}>0$. The spectral expansion
of $P$ can be written as

\[
P(x_{f},t_{f}|x_{0},0)=\frac{e^{-\beta U(x)}}{Z}+\sum_{\alpha\neq0}e^{-t_{f}E_{\alpha}}P_{\alpha}(x_{f},x_{0})\]

We see that for large $t_{f}$ the system converges to the Boltzmann
distribution, and that its relaxation time is given by the inverse
of the first eigenvalue $\tau_{R}=1/E_{1}$. In systems with high
energy barriers, such as proteins, the gap $E_{1}$ may be very small,
and consequently the time $\tau_{R}$ which in this case is identified
with the folding time, can be very long.

Using the Feynman path integral representation, we may thus write eq.(\ref{eq:matrix}) as
\cite{12}
 \begin{equation}
P(x_{f},t_{f}|x_{0},0)=e^{-\beta(U(x_{f})-U(x_{0}))/2}\int_{(x_{0},0)}^{(x_{f},t_{f})}\mathcal{D}x(t)\exp\left(-\frac{1}{4k_{B}T}\int_{0}^{t_{f}}dt\left(\gamma\dot{x^{2}}+\frac{1}{\gamma}V(x)\right)\right)\label{PI1}
\end{equation}

\begin{center}
\begin{figure}
\includegraphics[clip,width=15cm]{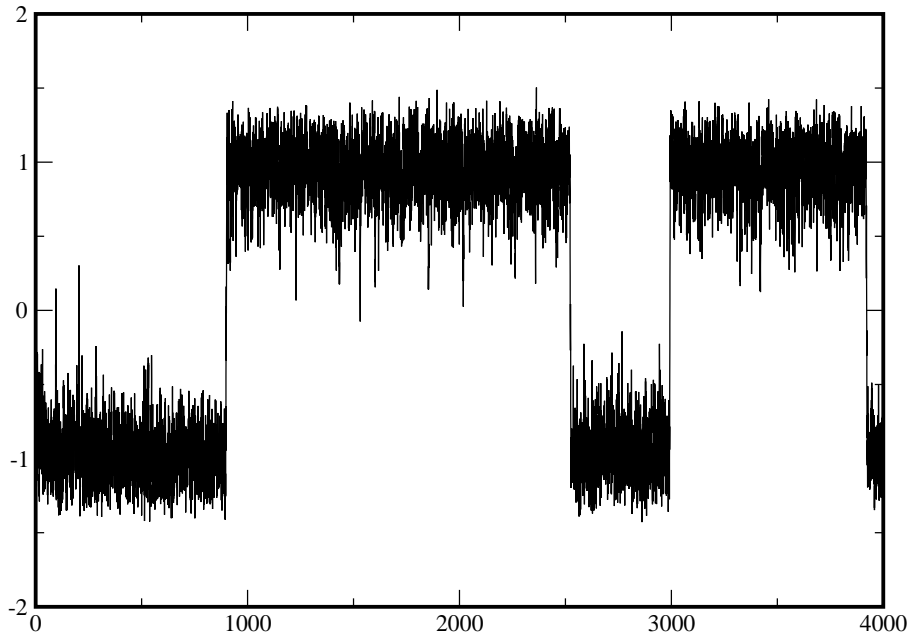} \caption{A long Langevin trajectory in the double-well. \label{fig1}}
\end{figure}
\end{center}

In the following, we will be mostly interested in problems of energy
or entropy barrier crossing, which are of utmost importance in many
chemical, biochemical or biological reactions. As we already mentioned
before, the archetype of such reactions is protein folding, a model
we will use in the rest of this paper. A protein is a small biopolymer,
which essentially may exist in two states, namely the native state
(with biological activity) and the denatured state (with no biological
activity) \cite{13}. The protein being a small system (up to a few
hundred amino-acids), it never stays in one of the two states, but
rather makes rare stochastic transitions between the two states 
(see fig.\ref{fig1}). The picture which emerges is that of the system staying
for a long time in one of the two states and then making a rapid transition
to the other state.

It follows that for most of the trajectory, the system makes uninteresting
stochastic oscillations in the well, and can be described by normal
mode analysis. Rarely, there is a very short but interesting physical
phenomenon, which is the fast transition from one minimum to the other.

This picture has been confirmed by single molecule experiments \cite{3,4},
where the waiting time in one state can be measured, but the time
for crossing from one state to the other is so short that it cannot
be resolved. This scenario has also been confirmed recently by very
long millisecond molecular dynamics simulations \cite{14} which for
the first time show spontaneous thermal folding-unfolding events.

According to Kramers theory, the total transition time $\tau_{K}$
(waiting + crossing) scales like the exponential of the barrier energy

\[
\tau_{K}\sim e^{\Delta E/k_{B}T}\]

The {}``Kramers time'' $\tau_{K}$ is the sum of two times: 
\begin{itemize}
\item the waiting time in the potential well 
\item the crossing time over the barrier $\tau_{C}$ 
\end{itemize}
It is well known that the crossing time $\tau_{C}$ is small compared
to $\tau_{K}$ and indeed, Hummer \cite{15} and subsequently Szabo \cite{16}
have shown that

\[
\tau_{C}\sim\ln\frac{\Delta E}{k_{B}T}<<\tau_{K}\]

These Kramers and crossing times are averages. In fact, these times
are distributed (random variables) and single molecule experiments
or long molecular dynamics simulations allow to compute their probability
distributions.

However, it seems a bit wasteful to simulate proteins over huge time
scales (milliseconds), during which only small conformational vibrations
occur, just to observe interesting physical crossing events which
occur very rarely, on the sub-microsecond scale.

The goal of this paper is to show how one can generate a representative
sample of transition paths, starting in state A at time $0$ and ending
in state B at some arbitrary time $t_{f}$. The typical times of interest
are not the (long) folding times, but rather the (very short) transition
or barrier crossing times. In mathematical terms, we are looking for
the paths starting from A at time $0$ and conditioned to end in state
B at time $t_{f}<<\tau_{K}$.

\section{The conditional probability}

Using the path integral representation of eq.(\ref{PI1}), we see that
the probability for a path $\{x(t)\}$ starting at $x_{0}$ at time
$0$, to end at $x_{f}$ at $t_{f}$ is given by

\begin{equation}
P(\{x(t)\})=\frac{1}{A}e^{-\beta(U(x_{f})-U(x_{0}))/2}\exp\left(-\frac{1}{4k_{B}T}\int_{0}^{t_{f}}dt\left(\gamma\dot{x^{2}}+\frac{1}{\gamma}V(x)\right)\right)\label{eq:pro1}\end{equation}
 where

\begin{equation}
A=\int dx_{f}e^{-\beta(U(x_{f})-U(x_{0}))/2}\int_{(x_{0},0)}^{(x_{f},t_{f})}\mathcal{D}x(t)\exp\left(-\frac{1}{4k_{B}T}\int_{0}^{t_{f}}dt\left(\gamma\dot{x^{2}}+\frac{1}{\gamma}V(x)\right)\right)\label{eq:pro2}\end{equation}

The conditional probability over all paths starting at $x_{0}$ at
time $0$ and ending at $x_{f}$ at time $t_{f}$, to find the system
at point $x$ at an intermediate time $t$ is given by\[
\mathcal{P}(x,t)=\frac{1}{P(x_{f},t_{f}|x_{0},0)}Q(x,t)P(x,t)
\label{condition}
\]
 where

\[
P(x,t)=P(x,t|x_{0},0)\]

\[
Q(x,t)=P(x_{f},t_{f}|x,t)\]

The equation satisfied by $P$ is given by (\ref{eq:FP}), whereas
that for $Q$ is given by

\begin{equation}
\frac{\partial Q}{\partial t}=-D\frac{\partial^{2}Q}{\partial x^{2}}+D\beta\frac{\partial U}{\partial x}\frac{\partial Q}{\partial x}\label{eq:FTadj}\end{equation}

It follows easily that the equation for the conditional probability
$\mathcal{P}(x,t)$ is given by

\[
\frac{\partial\mathcal{P}}{\partial t}=D\frac{\partial}{\partial x}\left(\frac{\partial\mathcal{P}}{\partial x}+\frac{\partial}{\partial x}\left(\beta U-2\ln Q\right)\mathcal{P}\right)\]

Comparing this equation with the initial Fokker-Planck (\ref{eq:FP})
and Langevin (\ref{eq:langevin}) equations, one sees that it can
be obtained from a Langevin equation with a modified potential

\begin{equation}
\frac{dx}{dt}=-\frac{D}{k_{B}T}\frac{\partial U}{\partial x}+2D\frac{\partial\ln Q}{\partial x}+\eta(t)\label{eq:bridge1}\end{equation}

This equation has been previously obtained using the Doob transform
\cite{17} and is known in the probability literature as a \emph{Langevin
bridge}: the paths $\{x(t)\}$ generated by (\ref{eq:bridge1}) are
conditioned to end at $(x_{f},t_{f}).$
It is the new term in the Langevin equation that guarantees that the
trajectory starting at $(x_{0},0)$ will end at $(x_{f},t_{f}).$

Using eq.(\ref{eq:matrix}) for $Q$, one can write equation (\ref{eq:bridge1})
as

\begin{equation}
\frac{dx}{dt}=2\frac{k_{B}T}{\gamma}\frac{\partial}{\partial x}\ln<x_{f}|e^{-(t_{f}-t)H}|x>+\eta(t)\label{eq:bridge2}\end{equation}

Using the analogous of the correspondence principle of quantum mechanics
\cite{18}, i.e. $\frac{\hbar}{i}\frac{\partial}{\partial x}\rightarrow p$,
this equation can also be rewritten in the form 
\begin{equation}
\frac{dx}{dt}=<\dot{x}(t)>+\eta(t)\label{eq:bridge3}
\end{equation}
 where by definition 
\begin{equation}
<\dot{x}>=\frac{1}{<x_{f}|e^{-(t_{f}-t)H}|x>}\int_{(x,t)}^{(x_{f},t_{f})}\mathcal{D}x(\tau){\dot{x}(t)}\exp\left(-\frac{1}{4k_{B}T}\int_{t}^{t_{f}}d\tau\left(\gamma\dot{x^{2}}+\frac{1}{\gamma}V(x)\right)\right)\end{equation}
 Note that for large time $t_{f}$, the matrix element in eq.(\ref{eq:bridge2})
is dominated by the ground state of $H$, namely $<x_{f}|e^{-(t_{f}-t)H}|x>\sim e^{-\frac{\beta}{2}(U(x_{f})+U(x))}$
and as expected one recovers the standard (unconditioned) Langevin
equation.

Since we have a natural splitting of the Hamiltonian $H$ as $H=H_{0}+V_{1}$
with $H_{0}=-\frac{k_{B}T}{\gamma}\frac{\partial^{2}}{\partial x^{2}}$
and $V_{1}=V/4\gamma k_{B}T$, it is convenient to rewrite the above
equation as

\begin{equation}
\frac{dx}{dt}=2\frac{k_{B}T}{\gamma}\frac{\partial}{\partial x}\ln<x_{f}|e^{-(t_{f}-t)H_{0}}|x>+2\frac{k_{B}T}{\gamma}\frac{\partial}{\partial x}\ln\frac{<x_{f}|e^{-(t_{f}-t)H}|x>}{<x_{f}|e^{-(t_{f}-t)H_{0}}|x>}+\eta(t)\label{eq:bridge4}\end{equation}

Note that the first term in the r.h.s. above is singular at $t=t_{f}$
and is thus responsible for driving the system to $(x_{f},t_{f})$
whereas the second one is regular. It follows that the first term
is the only term which can drive the system to $(x_{f},t_{f})$, and
any approximation which keeps the second term finite for $t=t_{f}$
will not affect this property.

This nice bridge equation cannot be used \textquotedbl{}as is\textquotedbl{},
since we don't know how to compute the function $Q$ or equivalently
the matrix element in the above equation. There are many ways to approximate
this function. It is important however, to preserve detailed balance as well as possible, that the approximation retains
the symmetry of the matrix element.

\section{The modified Langevin equation and reweighting}
The only approximation we found which remains local in time, i.e.
which does not give rise to an integro-differential stochastic equation
is the symmetric form of the Trotter approximation, commonly used
in quantum mechanics \cite{12}. Indeed, for short times $t$,
a very simple and convenient symmetric approximation for $Q$ is given
by

\begin{equation}
e^{-Ht}\sim e^{-tV_{1}/2}e^{-tH_{0}}e^{-tV_{1}/2}+O(t^{3})\label{eq:trotter}\end{equation}
 which translates into

\[
<x_{f}|e^{-Ht}|x>\sim e^{-\frac{\beta\gamma}{4t}(x_{f}-x)^{2}-\frac{\beta t}{8\gamma}(V(x_{f})+V(x))}\]

It would be nice to relate the range of validity of this equation
to the spectrum of $H$. Indeed, as was shown before, the spectrum
of $H$ corresponds to all the dynamical times of the system (folding
times, transition times, etc...). We have not succeeded in finding
such a relation except in the solvable case of the harmonic oscillator.
In that case, it can easily be shawn that the natural expansion parameter
is $t\Delta$ where $\Delta$ is the constant gap between the energy
levels of $H$. As mentioned before, in the case of protein folding,
the folding time which is the inverse of the first gap of the system
can be very long, and we might expect the above approximation to be
valid for times much smaller than this time. In particular, this approximation
would allow to investigate the crossing times, much shorter than
the folding time.

Plugging eq.(\ref{eq:trotter}) in eq.(\ref{eq:newbridge}) we
obtain the approximate Langevin bridge equation which in arbitrary
dimension (or with arbitrary number of degrees of freedom) reads

\begin{equation}
\frac{d\vec{x}}{dt}=\frac{\vec{x}_{f}-\vec{x}}{t_{f}-t}-\frac{1}{4\gamma^{2}}(t_{f}-t)\nabla V(\vec{x})+\vec{\eta}(t)\label{eq:approx}\end{equation}
 where $\vec{\eta}(t)$ is a white noise vector whose components satisfy
the relations (\ref{eq:noise1}) and (\ref{eq:noise2}) and

\begin{equation}
V(\vec{x})=\left(\nabla U\right)^{2}-2k_{B}T\nabla^{2}U\label{eq:V}\end{equation}

The first term in the r.h.s of eq.(\ref{eq:approx}) is the one which
drives the particle to reach $x_{f}$ at time $t_{f}$.
The potential which governs this bridge equation is not the original
$U(x)$ but rather the effective potential $V(x)$. Note also that
the force term is proportional to $(t_{f}-t)$ and thus becomes small
as the particle gets close to its target site.

In order to build a representative sample of paths starting at $(x_{0},0)$
and ending at $(x_{f},t_{f}),$ one must simply solve this equation
for many different realizations of the random noise. 
Only the initial boundary condition is to be imposed,
as the singular term in the equation imposes the correct final boundary
condition. An important point to note is that all the trajectories
generated by eq.(\ref{eq:approx}) are statistically independent.
From a numerical point of view, this means that this equation can
be fully parallelized, and from a statistical point of view, it implies
that all trajectories can be used in the representative sample. This
last important point is to be contrasted with most existing methods
where the sample are generated by some stochastic (Monte Carlo) methods
which generate highly correlated trajectories.

Before presenting examples of application of this method, let us discuss
how to correct for the fact that the total time $t_{f}$ should be
small for the approximation to be valid.

Due to this restriction, the statistic of trajectories is not exact
for larger times. Indeed, if eq.(\ref{eq:approx}) were exact, each
trajectory would be generated with its correct weight, and if one
wanted to calculate observables, one would just have to compute simple
white averages over all trajectories. However, as the equation is approximate,
one needs to resample the ensemble of trajectories, that is, assign
them a new weight. As we will show, the resampling weight is easily
obtained.

Indeed, if we consider the sample of trajectories generated
using eq.(\ref{eq:approx}) between $(x_{0},0)$ and $(x_{f},t_{f})$,
the weight of each trajectory should be given by eq.(\ref{eq:pro1}).
However, it is clear from eq.(\ref{eq:approx}) that, using the Ito
prescription, the weight with which it was generated is given by

\begin{equation}
\exp\left(-\frac{\gamma}{4k_{B}T}\int_{0}^{t_{f}}dt\left(\frac{d\vec{x}}{dt}-\frac{\vec{x}_{f}-\vec{x}}{t_{f}-t}+\frac{t_{f}-t}{4\gamma^{2}}\nabla V(\vec{x})\right)^{2}\right)\label{eq:weight1}\end{equation}

Up to a normalization, the reweighting factor for a trajectory is
thus given by

\begin{equation}
\exp\left(-\frac{\gamma}{4k_{B}T}\int_{0}^{t_{f}}dt\left(\left(\frac{d\vec{x}}{dt}+\frac{1}{\gamma}\nabla U\right)^{2}-\left(\frac{d\vec{x}}{dt}-\frac{\vec{x}_{f}-\vec{x}}{t_{f}-t}+\frac{t_{f}-t}{4\gamma^{2}}\nabla V(\vec{x})\right)^{2}\right)\right)\label{eq:reweight}\end{equation}

This quantity is easily calculated and allows for a correct evaluation
of averages over paths.

This reweighting technique can also be used to generate paths
statistically exactly sampled according to 
eq.(\ref{eq:bridge1}).
Indeed, consider eq.(\ref{eq:bridge3}). The expectation value $<\dot{x}(t)>$
can be computed by generating at each time $t$ an ensemble of (approximate)
trajectories starting from the current point $x$ at time $t$ and ending at
$x_{f}$ at time $t_{f}$ by using eq.(\ref{eq:approx}). By  reweighting
them using the weights of eq.(\ref{eq:reweight}), we can reliably compute $<\dot x(t)>$ and thus solve eq.(\ref{eq:bridge3}). 
Note that this procedure
which generates correctly weighted trajectories might seem computationally
costly. However, since all trajectories are
independent, they can efficiently be generated using massive parallelization.

\section{The native state}

Eq.(\ref{eq:trotter}) is in fact not quite valid between non normalizable
states like $|x>$ and $|x_{f}>$, in that it is not true to order
$O(t^{3})$. However it is true between a normalizable state and $|x>$.
Assume that the final state of the system is defined by a probability
distribution $\phi(x)$. For instance, for the case of a protein,
$\phi(x)$ could represent the Boltzmann weight around the native
state of the protein. The probability for the system to start at $x$
at time $t$ and end at time $t_{f}$ in the native state is given
by

\begin{equation}
Q(x,t)=\int dy\ \phi(y)P(y,t_{f}|x,t)\label{eq:native}
\end{equation}
or using (\ref{eq:matrix})
\begin{equation}
Q(x,t)=\int dy\ \phi(y)e^{-\beta(U(y)-U(x))/2}<y|e^{-(t_f - t) H}|x>\label{eq:native2}\end{equation}
 where $\phi$ restricts the integration over $y$ to the vicinity
of the native state.

With this definition of $Q$, it is straightforward to see that 
eq.(\ref{eq:FTadj})
and (\ref{eq:bridge1}) are still valid.

Using the approximation (\ref{eq:trotter}) we can write

\begin{equation}
Q(x,t)=e^{\beta U(x)/2}e^{-\frac{t}{2}V_{1}(x)}\int\frac{dy}{A}\ \phi(y)e^{-\beta U(y)/2}e^{-\frac{t}{2}V_{1}(y)}e^{-\frac{\beta\gamma}{4}\frac{(y-x)^{2}}{t_{f}-t}}\label{eq:native3}
\end{equation}
where $A=\sqrt{4\pi(t_{f}-t)/\beta\gamma}$.

As the function $\phi$ restricts the integration in (\ref{eq:native3})
to the vicinity of the native state, we can approximate the potential
$U(x)$ in this region by a quadratic expansion in terms of the normal
modes

\[
U(x)\simeq\frac{\omega}{2}(x-x_{f})^{2}\]

It follows that $V$ and $V_{1}$ are also quadratic and thus the
integral (\ref{eq:native3}) can be performed. Although we will consider
only one-dimensional cases in the examples, we present the results
for the multi-dimensional case.

Denoting by $\omega_{ij}=\frac{\partial^{2}U}{\partial x_{i}\partial x_{j}}|_{x_{i}^{f}}$
the Hessian matrix of normal modes around the native state, the potential $U$ can
be written in that region as

\[
U(x)=\frac{1}{2}\sum_{i,j}(x_{i}-x_{i}^{f})\omega_{ij}(x_{j}-x_{j}^{f})\]
 and thus

\[
V(x)=\sum_{i,j}(x_{i}-x_{i}^{f})\Omega_{ij}(x_{j}-x_{j}^{f})-2k_{B}T{\rm \, Tr}\ \omega_{ij}\]
 where the symbol Tr denotes the trace of the normal mode matrix and

\[
\Omega_{ij}=\sum_{k} \omega_{ik}\omega_{kj}\]

The function $Q$ can be easily calculated as

\[
Q(x,t)=e^{-\frac{\beta}{8\gamma}(t_{f}-t)V(x)-\frac{\beta}{4}\sum_{i,j}(x_{I}^{f}-x_{i})W_{ij}(x_{j}^{f}-x_{j})}\]
 where

\[
W_{ij}=\sum_{k}D_{ik}(I+\frac{t_{f}-t}{\gamma}D)_{kj}^{-1}\]
where $I$ is the unit matrix and
\[
D_{ij}=\omega_{ij}+\frac{t_{f}-t}{2\gamma}\Omega_{ij}\]

The bridge equation becomes then

\begin{equation}
\frac{dx_{i}}{dt}=\frac{1}{\gamma}\sum_{j}W_{ij}(x_{j}^{f}-x_{j})-\frac{1}{4\gamma^{2}}(t_{f}-t)\nabla_{i}V(\vec{x})+\eta_{i}(t)\label{eq:newbridge}\end{equation}

\section{Including the solvent}
In many cases, in particular protein folding, one wants to include explicitly the solvent molecules, most often water. It is thus desirable to generate trajectories which are conditioned for the protein coordinates, but not for the water molecules.

Denoting by $X_i$ the set of coordinates of the water molecules, the conditional probability, over all paths starting at $\{x_0, X_0\}$ at time 0 and ending at  $x_f$ at time $t_f$ (irrespective of the position of the solvent molecules), for the system to be at $\{x,X\}$ at time $t$ is given by

\[
\mathcal{P}(x,X,t)=\frac{1}{\int dX_f P(x_{f},X_f,t_{f}|x_{0},X_0,0)}Q(x,X,t)P(x,X,t)
\label{condition1}
\]
 where

\begin{eqnarray}
P(x,X,t)&=&P(x,X,t|x_{0},X_0,0) \\
Q(x,X,t)&=& \int dX_f P(x_{f},X_f,t_{f}|x,X,t) \label{Q}
\end{eqnarray}

The coordinates $X_f$ of the solvent are integrated over since the trajectories are not conditioned over the solvent molecules.

Using the method described in the previous sections, the exact generalized Langevin equations satisfied by the coordinates are

\begin{eqnarray}
\frac{dx_i}{dt} &=&-\frac{D_1}{k_{B}T}\frac{\partial U}{\partial x_i}+2D_1\frac{\partial\ln Q}{\partial x_i}+\eta^{(1)}_i(t)  \\
\frac{dX_i}{dt}&=&-\frac{D_2}{k_{B}T}\frac{\partial U}{\partial X_i}+2D_2\frac{\partial\ln Q}{\partial X_i}+\eta^{(2)}_i(t) \label{solvent}
\end{eqnarray}
where $D_1$ and $D_2$ are resp. the diffusion constants for protein and water molecules and the Gaussian noises $\eta^{(1,2)}_i(t)$ satisfy the relation

\begin{equation}
<\eta^{(1,2)}_i(t)\ \eta^{(1,2)}_j(t')> = 2 D_{1,2} \delta_{ij} \delta (t-t') \label{noise}
\end{equation}

Consider eq.(\ref{Q}). Let us show that the additional force term vanishes. Indeed,
\begin{eqnarray}
Q(x,X,t)&=&\int dX_f P(x_{f},X_f,t_{f}|x,X,t) \nonumber \\
&=& \int dX_f P(x_f-x,X_f-X,t_f-t) \label{integral}
\end{eqnarray}
because of space and time translation invariance.
Due to the integration over $X_f$ in eq.(\ref{integral}), we see that $Q(x,X,t)$ does not depend on $X$, and thus the new drift term in (\ref{solvent}) is absent. 
Therefore the exact equations for the conditional probability in presence of solvent are
\begin{eqnarray}
\frac{dx_i}{dt} &=&-\frac{D_1}{k_{B}T}\frac{\partial U}{\partial x_i}+2D_1\frac{\partial\ln Q}{\partial x_i}+\eta^{(1)}_i(t)  \\
\frac{dX_i}{dt}&=&-\frac{D_2}{k_{B}T}\frac{\partial U}{\partial X_i}+\eta^{(2)}_i(t)
\end{eqnarray}

Using the Trotter approximation (\ref{eq:trotter}), these equations become (using vector notations)
\begin{eqnarray}
\frac{d\vec{x}}{dt}&=&\frac{\vec{x}_{f}-\vec{x}}{t_{f}-t}-\frac{1}{4\gamma^{2}}(t_{f}-t)\nabla_x V(\vec{x},\vec{X})+\vec{\eta^{(1)}}(t) \\
\frac{d\vec{X}}{dt}&=&-\frac{D_2}{k_{B}T}\frac{\partial U(\vec{x}, \vec{X})}{\partial \vec{X}}+\vec{\eta^{(2)}}(t)
\end{eqnarray}
where the noises are Gaussian, correlated according to eq.(\ref{noise}).

We thus conclude that in presence of the solvent, the protein is evolved through a modified Langevin equation with the effective potential $V(\vec x,\vec X)$, whereas the solvent molecules are evolved according to the standard Langevin equation in presence of the original potential $U(\vec x, \vec X)$. 

Extension of this method to the case of the native state (see previous section) is immediate.

\section{Example: The quartic double-well}

We now illustrate the method on the example of barrier crossing in
1d (quartic potential).

\[
U(x)=\frac{1}{4}(x^{2}-1)^{2}\]

\begin{figure}
\includegraphics[clip,width=15cm]{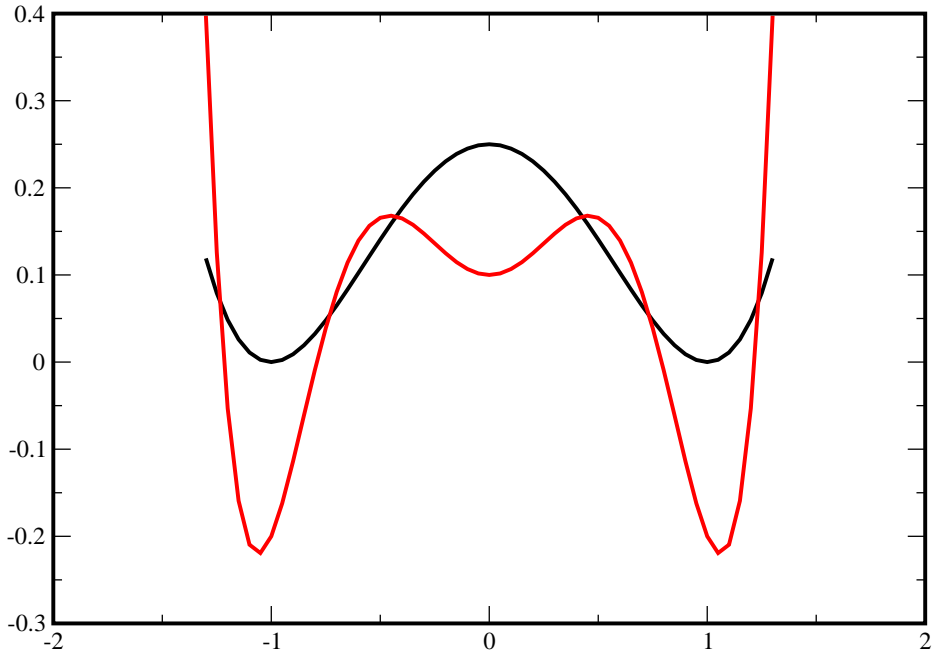}\caption{Potential $U(x)$ (in black) and potential $V(x)$ (in red). \label{potential}}

\end{figure}

This potential has two minima at $x=\pm1$, separated by a barrier
of height 1/4. Note that at low enough temperature, the potential
$V(x)$ has two minima at points close to $\pm1$ and one minimum
at $x=0$ (from eq.(\ref{potential})). Note that $V(x)$
is much steeper than $U(x)$ and thus more confining, around its minima.

The model can be solved exactly by solving numerically the Fokker-Planck
equation or by diagonalizing the Hamiltonian. All the examples are performed
at low temperature $T=0.05$, where the barrier height is equal to
5 in units of $k_{B}T$ and the Kramers relaxation time, given by
the inverse of the smallest non-zero eigenvalue of $H$, is equal
to $\tau_{K}=366.39$.

\begin{figure}
\includegraphics[clip,width=15cm]{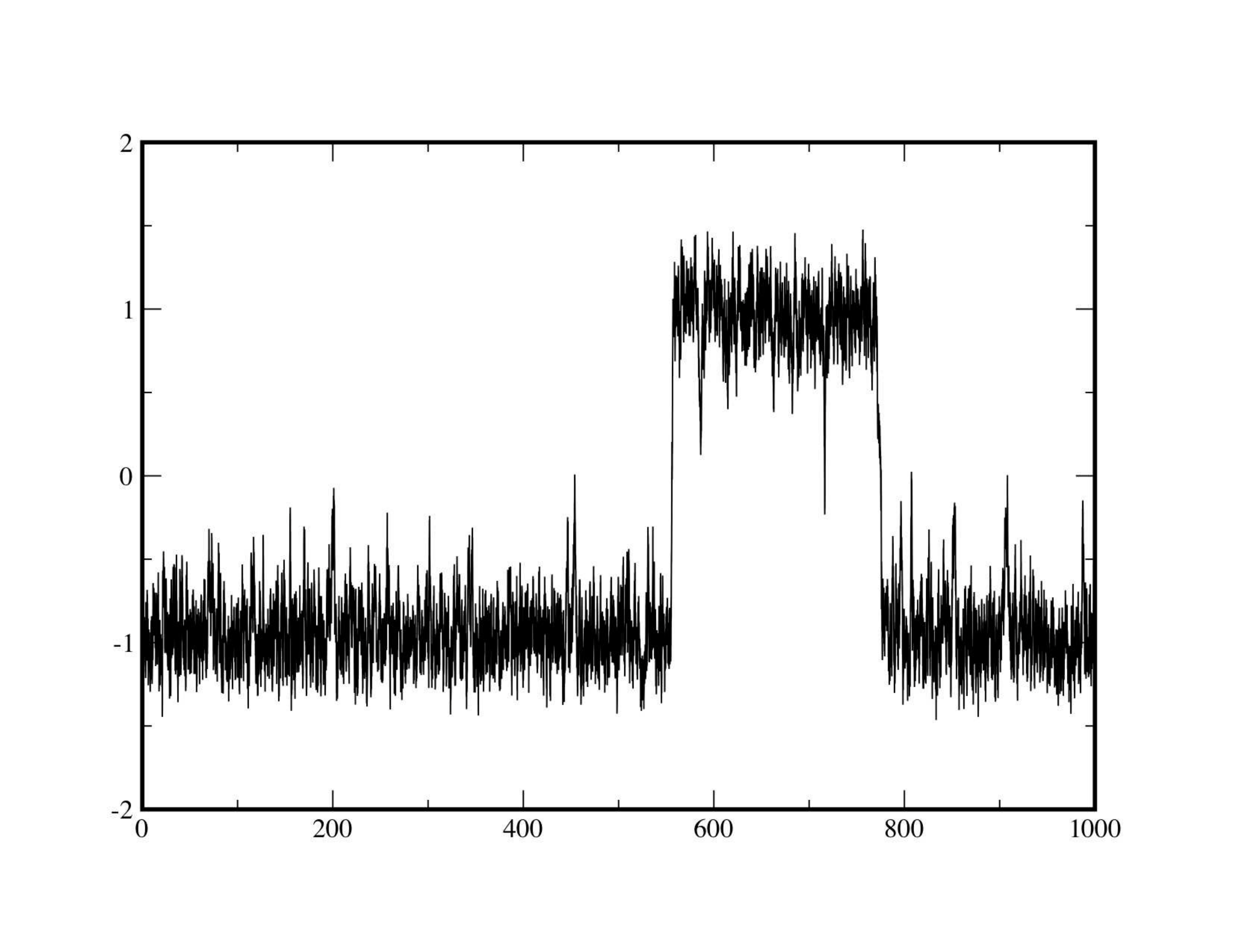}\caption{Full Langevin trajectory during time $t_{f}=1000$ with 2 transitions
between the minima\label{fig:Full-Langevin-trajectory}}
\end{figure}

On fig.\ref{fig:Full-Langevin-trajectory}, we present a long trajectory
($t_{f}=1000)$ obtained by solving the Langevin eq.(\ref{eq:langevin})
for a particle starting at $x_{0}=-1$ at time 0. The general pattern
described in the introduction can be easily checked: the particle
stays in the left well for a time of the order of 550, then jumps
very rapidly into the right well, where it stays for a time of the
order of 200, then jumps back to the left well where it stays again
a time equal to about 250.

\begin{figure}
\includegraphics[clip,width=15cm]{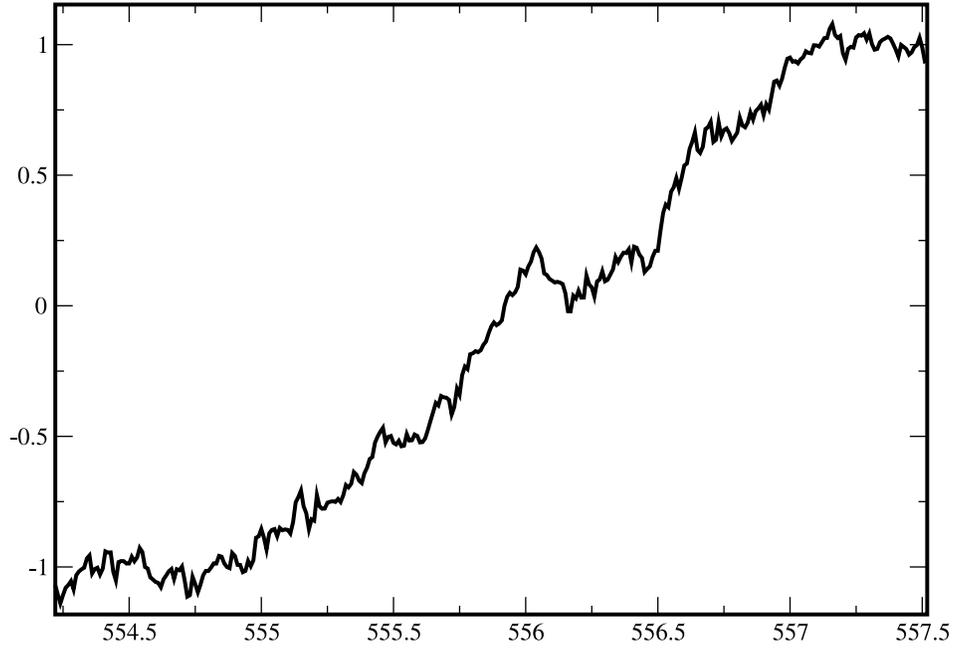}\caption{Enlargement of the first transition region\label{fig:Enlargement-of-the}}
\end{figure}

The two crossings times are very short, and we display
an enlargement of the first transition in fig.\ref{fig:Enlargement-of-the}.

As can be seen, the crossing time for this specific trajectory is approximately  $\tau_{C}\thickapprox2.5$,
much smaller than the Kramers time.

In fig.\ref{fig:another-set-of}, we plot two examples of two trajectories
conditioned to cross the barrier during a time $t_{f}=5.$ The trajectory
in black is obtained by solving the exact bridge eq.(\ref{eq:bridge2})
by computing exactly (using a spectral decomposition) the
matrix element of the evolution operator, while the trajectory in
red is obtained by solving the approximate eq.(\ref{eq:approx})
with the exact same sequence of noise $\eta(t)$. In the left figure,
the 2 trajectories are barely distinguishable, whereas the agreement
is not as spectacular on the right figure.

\begin{center}
\begin{figure}
\includegraphics[clip,width=7cm]{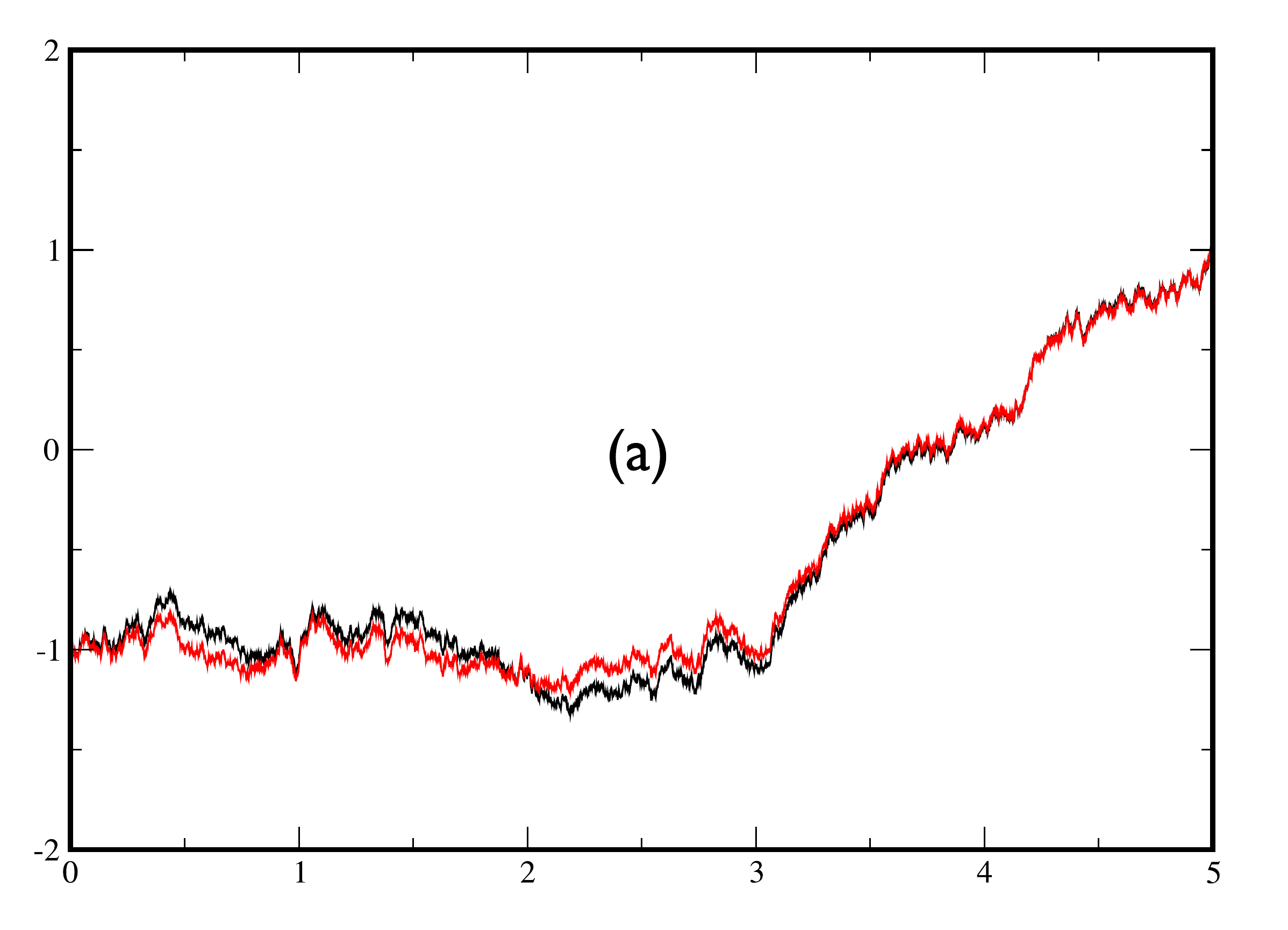} \includegraphics[clip,width=7cm]{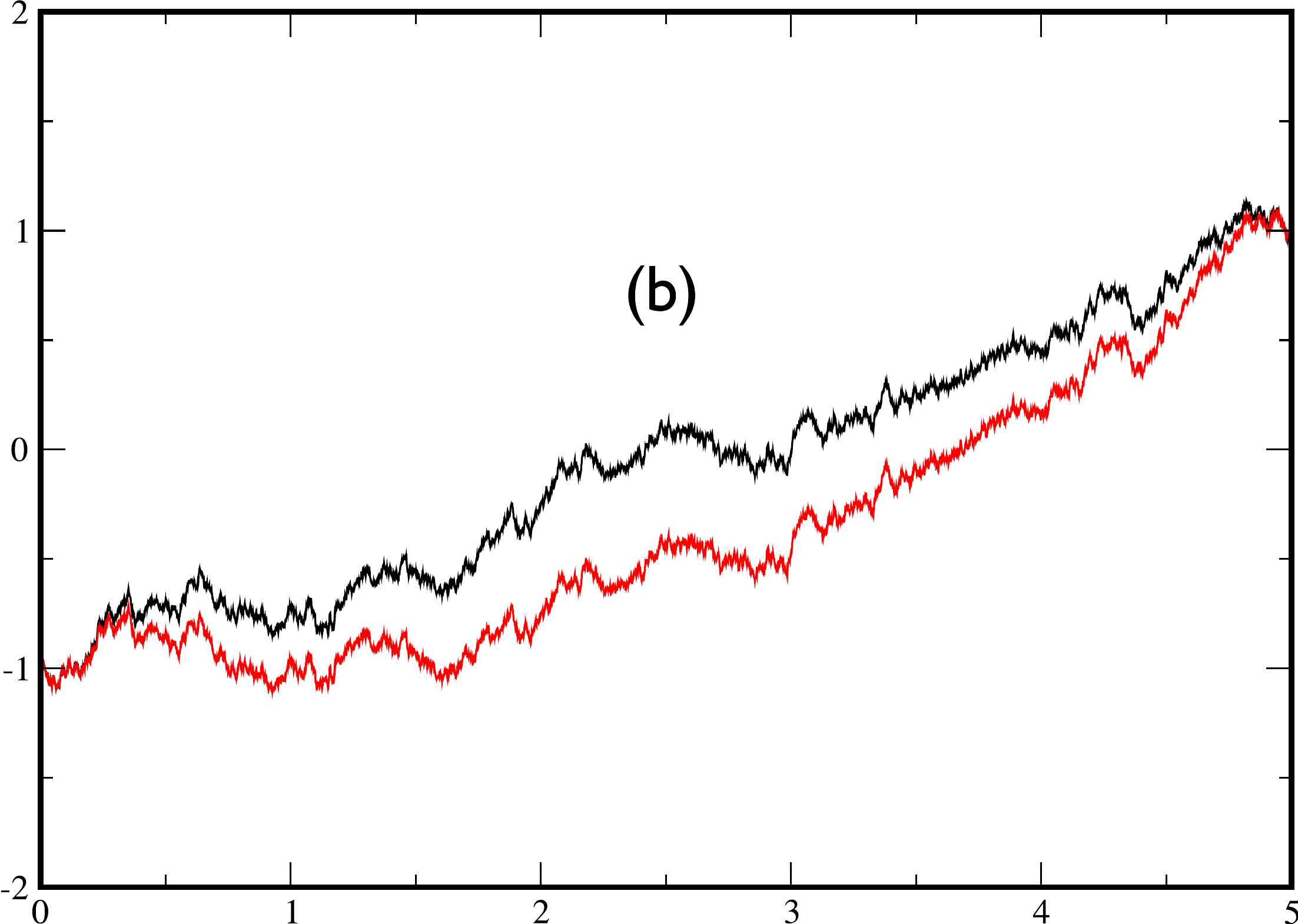}
\caption{Two sets (a) and (b) of exact trajectories (in black) and approximate
trajectories (in red)\label{fig:another-set-of}}

\end{figure}

\par\end{center}

Next we look at some observables, obtained by averaging over many
trajectories.

\begin{figure}
\includegraphics[clip,width=5cm]{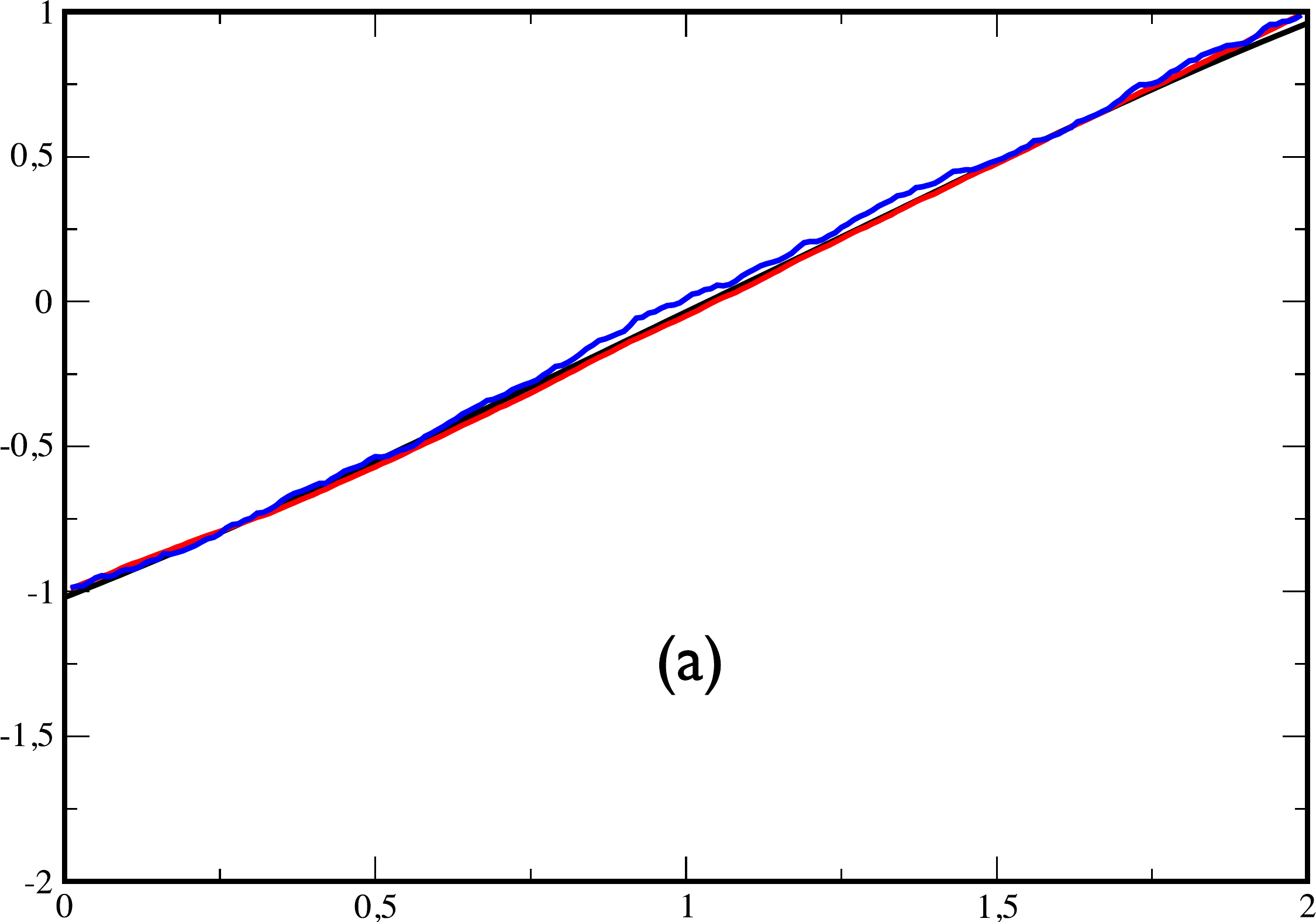} \includegraphics[clip,width=5cm]{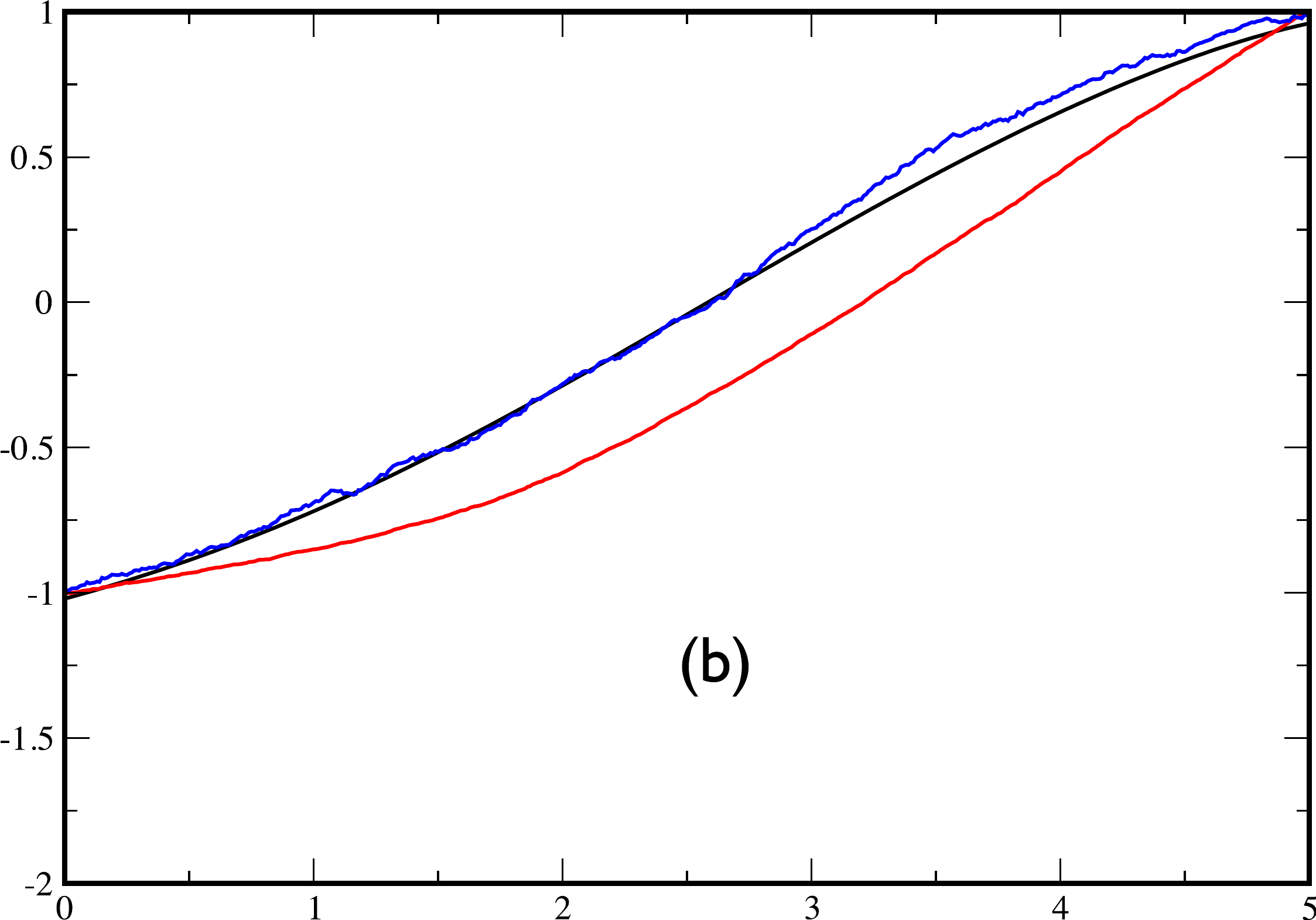}
\includegraphics[clip,width=5.1cm]{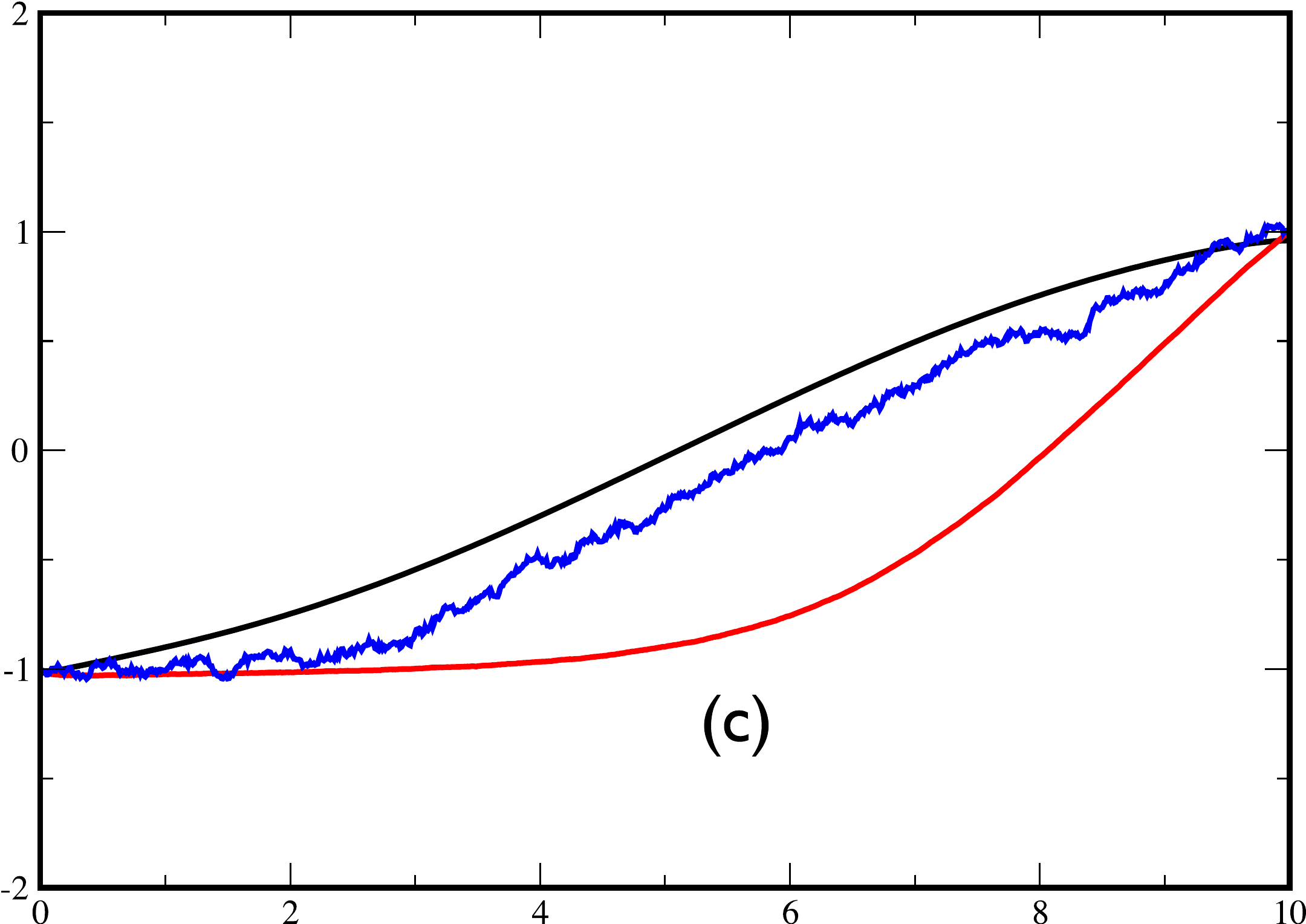} \caption{Average position as a function of time for (a) $t_{f}=2$, (b) $t_{f}=5$,
(c) $t_{f}=10$. Black curve: exact. Red curve: approximate. Blue curve:
reweighted \label{f3} }
\end{figure}

In fig.\ref{f3}, we plot: in black the exact average $x(t)$ (obtained
by a full expansion over the eigenstates of $H$), in red the average
$x(t)$ over 2000 trajectories obtained by solving eq.(\ref{eq:approx}),
and in blue, the average $x(t)$ obtained by reweighting the
trajectories according to eq.(\ref{eq:reweight}) . Plot (a) is obtained
for $t_{f}=2$, plot (b) for $t_{f}=5$ and plot (c) for $t_{f}=10$.

%

As expected, we see that the discrepancy between the exact (black)
and the approximate (red) average $x(t)$ increases with $t_{f}$.
For times shorter than the transition time $\tau_{C}$, the agreement
is excellent, whereas for $t_{f}=10>\tau_{C}$, the agreement is not
as good. However, we see that the reweighting procedure, although
not perfect, improves drastically the quality of the average for large
$t_{f}$.

\begin{figure}
\includegraphics[clip,width=8cm]{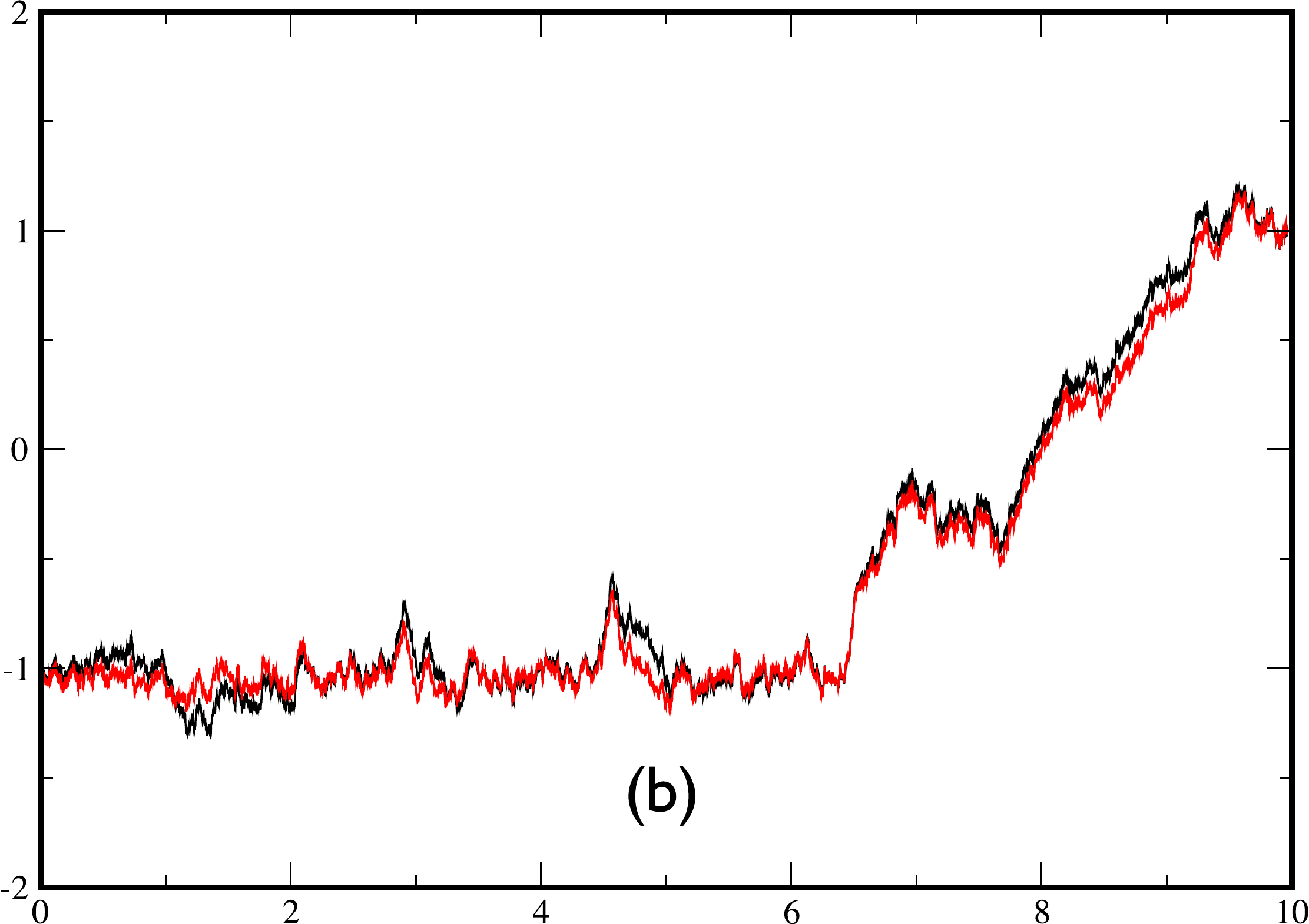} \includegraphics[clip,width=8cm]{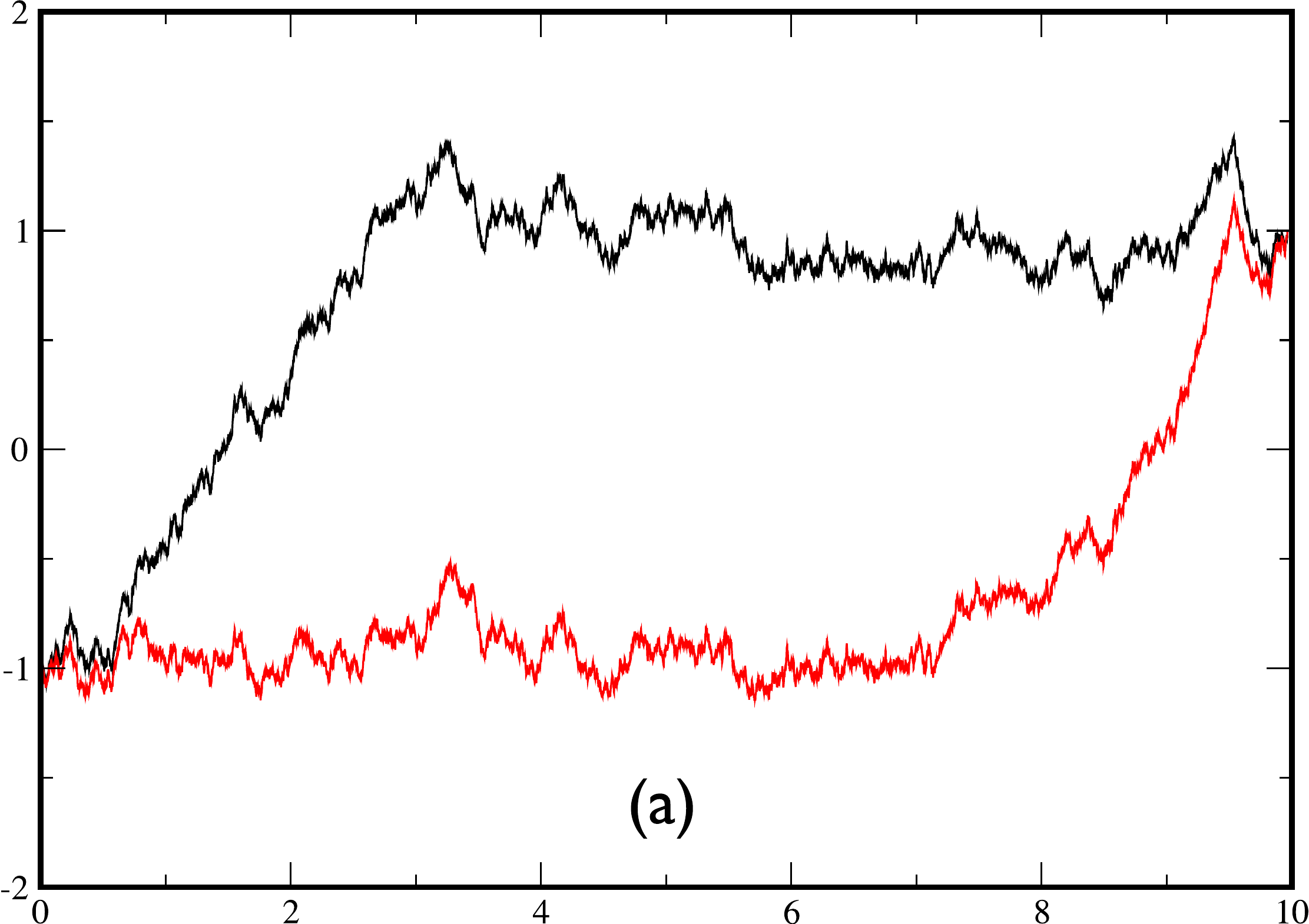}
\caption{Two sets (a) and (b) of trajectories conditioned to cross the barrier.
In black, exact trajectories and in red, approximate trajectories.
\label{f4}}
\end{figure}

One of the main defects which appears in the approximate theory is
the following: In the exact theory, the transition between the 2 minima
can take place at any time between 0 and $t_{f}$. By contrast, it
seems that in the approximate theory, the transition is driven by
the final state and takes place only in the end of the trajectory.
This effect remains negligible as long as $t_{f}\lesssim\tau_{C}$
but becomes important for $t_{f}>$$\tau_{c}$. We illustrate this
problem in fig.\ref{f4} for $t_{f}=10$. On the left figure, the
exact and approximate trajectories make their transition in the last
part of the time, whereas in the right figure, the real trajectory
crosses in the beginning while the approximate trajectory still crosses
in the last part.

However, as we are interested quantitatively only in the
region where the particle crosses the barrier, one can make long runs of approximate trajectories: They will
not be good approximations of the real trajectories, except in the
end of the trajectory where the transition to the final state occurs.

\section{Conclusion}

We have presented in this paper a novel method to generate paths following
the Langevin overdamped dynamics, starting from an initial configuration
and conditioned to end in a given final configuration (point or region
of configuration space). We propose an approximation which is valid
for small times. We have not been able to quantify how small should
the time be, but the approximate dynamics seems to correctly reproduce
the transition through a barrier. The approximate dynamics seems to
have a tendency to confine the system in its initial configuration,
and to allow for the transition only in the final stages. But this
is not really a drawback since if we evolve approximately the system
over long times, it will remain close to its initial condition, thus generating unreliable trajectories. However in the
latest stages, the system will make its transition to the final state
during a short time for which our approximation is reliable.

One of the great advantages of this method is that all generated trajectories
are statistically independent. It is thus very easy to generate many
of these trajectories using parallel computers. In addition, the trajectories
can be reweighted to provide a faithful sample of the exact stochastic
dynamics. Finally, this reweighting technique allows for the calculation
of the matrix element of the evolution operator, and thus 
allows for the generation of adequately sampled paths.

The paths generated by our method can also be used either as initial
paths to perform Monte Carlo transition path sampling, or as initial
conditions for path minimization to determine Dominant Folding Paths.

The method is as simple to implement as ordinary Langevin dynamics,
and its application to simple models of protein folding is currently
under way. 
\begin{acknowledgments}
The author wishes to thank M. Bauer and K. Mallick for very useful
discussions.
\end{acknowledgments}

\end{document}